\documentstyle[11pt,epsfig]{article} \textheight 700pt \textwidth
480pt \oddsidemargin 0pt \voffset -2.5cm
\title{\bf On the origins of galactic magnetic fields}
\author{A. Borzou\thanks{email:a.borzou@mail.sbu.ac.ir}, H. R. Sepangi\thanks{email:hr-sepangi@sbu.ac.ir},
R. Yousefi\thanks{email:raz.yousefi@mail.sbu.ac.ir} and A. H.
Ziaie\thanks{email:am.ziaie@mail.sbu.ac.ir},\,\,
\\ {\small Department of Physics, Shahid Beheshti University, Evin, Tehran 19839,
Iran}}
\begin{document}
\maketitle 

\begin{abstract}
We present a five dimensional unified theory of gravity and
electromagnetism which leads to modified Maxwell equations,
suggesting a new origin for galactic magnetic fields. It is shown
that a region with nonzero scalar curvature would amplify the
magnetic fields under certain conditions.
\vspace{10mm}\\
\end{abstract}
\section{Introduction}
A large number of astrophysical observations based on Faraday
rotation, Zeeman splitting, polarization of optical starlight and
synchrotron emission prove the existence of magnetic fields in
galaxies. A variety of such observations suggest that these magnetic
fields are present in all galaxies and galaxy clusters. The
properties of these magnetic fields are characterized by their huge
spatial coherence scale which is less than 1 Mpc and a modest
strength which is of the order of $(10^{-7}-10^{-5}G)$ \cite{01}.
Today, the best numbers obtained for the total strength of the
magnetic field in galactic disks give 6 - 7 $\mu G$ \cite{03,07}.
Also, extra-galactic magnetic fields with large scales which have
strengths of $ \sim \mu G $ have been detected by both synchrotron
emission and Faraday rotation in clusters like Coma Cluster.
Furthermore, beyond clusters, in sparser regions, astonishing
magnetic fields of $\sim 0.1 \mu G$ were measured \cite{08,09}. In
our galaxy, average magnetic fields of the interstellar medium of
order $\sim 5\mu G$ is detected \cite{10,11}.

Up until now the primordial field theory and dynamo theory have been
the main ingredients used to explain the origin of galactic magnetic
fields. In the former, galactic magnetic fields which we observe
today would be the relics of a coherent magnetic field existing in
the early universe before galaxy formation. One can imagine that in
a collapsing protogalaxy, the lines of force of the ambient magnetic
field (which tend to be frozen into the highly conductive gas) would
have been compressed by the gas motions associated with this
collapse, and from then on, the differential rotation of the galaxy
would have wrapped them up about its center. In the absence of any
other process, the galactic magnetic field would be wound up 50 to
100 times the present time, i.e., much more than indicated by
observations. The traditional way of resolving this inconsistency is
to take into account the magnetic diffusion. But if magnetic
diffusion is sufficiently present parallel to the galactic plane to
avoid a tight wind-up of magnetic field lines, it must also be true
perpendicular to the plane for field lines to diffuse out of the
disk. Under these conditions, galactic magnetic fields would have
completely decayed away by now, without ever reaching the observed
strengths of a few $ \mu G $ \cite{12,13,14,15}. Primordial models
are unrealistic since they also neglect deviations from axisymmetric
rotation of the gas, such as spiral density waves. Furthermore, such
models suffer from a fundamental problem; shear by differential
rotation increases the average field strength, not the magnetic
flux, while magnetic flux is lost into intergalactic space by
magnetic diffusion. Estimation of turbulent diffusion leads to a
decay time of the field of only $10^{8}$ yr, meaning that to have a
field amplification, one must seek another processes such as field
amplification by gas flows or a dynamo effect \cite{14,06}. In
dynamo theory, according to Faraday's law of induction, moving a
conductor through a magnetic field (seed field) induces a current
from which a new magnetic field can arise (Ampere's law). There are
different views on the origin of this seed field. One possibility is
due to astrophysical processes, such as Bierman battery effect in
which the magnetic field is produced by an electric current due to
the non-coinciding surfaces of constant pressure and density
\cite{03,04,05} and the other possibility can be cosmological
(primordial origin). A truly cosmological magnetic field is one that
cannot be associated with collapsing or virialized structures.
Cosmological magnetic fields can include those that exist prior to
the epoch of galaxy formation as well as those that are coherent on
scales greater than the scale of the largest known structures in the
Universe, i.e., $\geq$ 50 Mpc \cite{01}. If the last possibility was
true then all galaxies would have the same seed field but the field
strength in galaxies today would be different because this mechanism
depends on the galaxy characteristics such as age, structure, etc.
The magnitude of this seed field according to a calculation based on
the age of a typical galaxy must have been about $10^{-20} G$
\cite{06,02}. It seems as if the true combination of the flows of
the fields amplifies the original induced field which would then
lead to the field amplification that we measure in our observations.
In the dynamo model, it is assumed that large scale magnetic fields
observed in disk galaxies and spiral galaxies arise from the
combined action of differential rotation $(\omega)$ and helical
turbulence $(\alpha)$, a process known as the $\alpha\omega$ dynamo
wherein, new field is regenerated continuously by the combined
action of these two effects \cite{01}. In this model, the
small-scale cyclonic turbulent motions which have acquired a
preferred sense of rotation under the action of the Coriolis force
stretch and twist the magnetic field lines, imparting a net
rotation, whereby the magnetic field is created in the direction
perpendicular to the prevailing field which is known as the
``alpha-effect.'' Thus, in dynamo theory, the large-scale
differential rotation stretches magnetic field lines in the
azimuthal direction about the galactic center, while small-scale
cyclonic turbulent motions regenerate, via the alpha-effect, the
meridional component of the field from its azimuthal component. It
is the combination of these two complementary mechanisms that leads
to magnetic field amplification in galaxies \cite{12,16}.

There are, however, fundamental questions concerning the nature of
the dynamo. Magnetic fields are abundantly present in elliptical
galaxies and their presence is revealed through observations of
synchrotron emission. The standard $\alpha\omega$ dynamo however
does not explain the existence of magnetic fields in non-rotating or
slowly rotating systems such as elliptical galaxies and clusters. In
addition, the time scale for field amplification in the standard
$\alpha\omega$ dynamo could be too long to explain the fields
observed in very young galaxies \cite{01}. Even if the dynamos work
properly, they still need a small seed magnetic field. Therefore,
the conventional $\alpha \omega$ dynamo theory may not be able to
explain the amplification of the weak seed magnetic fields $\sim
10^{-18}G$ to micro Gauss strengths after $10^{9}$ years. The basic
difficulty is that the growth rate of a large scale dynamo is
typically some fraction $\xi$ of the angular velocity $\Omega$ of
the galaxy. Typical numbers are $\Omega=30 Gyr^{-1}$ and $\xi
\approx 0.1-0.5$. Even in the most optimistic case, the
amplification factor after $t =1 Gyr$ is just $exp(0.5 \Omega
t)\approx10^{6.5}$ \cite{17}.

In this work we present an alternative mechanism for explaining the
origin of galactic magnetic fields. This is based on the
applications of the results obtained from a fresh look at the
unification of gravity and electromagnetism. Such a unification, as
is well known, was introduced in the early part of the last century
by Kaluza and Klein. Over the years, a large number of works
centering around the original ideas of Kaluza and Klein have
appeared. Our unification scheme is no exception, but leads to
Maxwell equations which are modified relative to the standard
theory. Surprisingly, such modifications could be used to account
for the the origin of galactic magnetic fields for which, as of now,
no completely satisfactory explanation exists. We therefore start by
a brief introduction to the theory whose results are of great
importance in explaining this phenomena. For detailed account, the
reader should consult \cite{19}.
\section{Unification of gravity and electromagnetism} Let us begin
by considering a charged particle in the presence of gravitational
and electromagnetic fields in free fall with kinetic energy defined
as \cite{18}
\begin{eqnarray}
\frac{1}{2}mV^{2}=\frac{1}{2}m g_{\mu\nu}
\dot{x}^{\mu}\dot{x}^{\nu}-\frac{1}{2}qA_{\mu}\dot{x}^{\mu}+\phi',\label{eq13}
\end{eqnarray}
\begin{eqnarray}
V^{2}=\left(\frac{ds}{d\tau}\right)^{2}=g_{\mu\nu}\dot{x}^{\mu}\dot{x}^{\nu}-
\frac{q}{m}A_{\mu}\dot{x}^{\mu}+\phi.\label{eq14}
\end{eqnarray}
As for the metric describing the space-time, we write
\begin{equation}
ds^{2}=g_{\mu\nu}dx^{\mu}dx^{\nu}+f_{\mu}dx^{\mu}dl+\phi dl dl,
\end{equation}
and assume that $\frac{dl}{d\tau}=-1$, with
$f_{\mu}=\frac{q}{m}A_{\mu}$ being a vector and $\phi$ a scalar,
both in 4-dimensional space-time and
$\dot{x}^{\mu}=\frac{dx^{\mu}}{d\tau}$. Also, let us define
\begin{equation}
ds=(dx^{0},dx^{1},dx^{2},dx^{3},dl)\hspace{1cm}
\mbox{and}\hspace{1cm}V=(\dot{x}^{0},\dot{x}^{1},\dot{x}^{2},\dot{x}^{3},-1),
\end{equation}
and the five dimensional metric as
\begin{equation}
\hat{g}_{_{AB}}=\left(%
\begin{array}{ccccc}
  g_{\alpha\beta} &   &   &   & f_{0} \\
    &   &   &   & f_{1} \\
    &   &   &   & f_{2} \\
    &   &   &   & f_{3} \\
  f_{0} & f_{1} & f_{2} & f_{3} & \phi \\
\end{array}%
\right),
\end{equation}
We now assume that all geometrical objects that we define are
independent of the fifth coordinate and thus their derivatives with
respect to the fifth coordinate is zero. We may then obtain the
geodesic equation by minimizing the usual integral
\begin{equation}
\delta s=\delta\int{ds}=\delta\int{Vd\tau}=0,
\end{equation}
resulting in
\begin{equation}
\ddot{x}^{\kappa}+\frac{1}{2}g^{\lambda
\kappa}(g_{\lambda\beta,\alpha}+g_{\lambda\alpha,\beta}-
g_{\alpha\beta,\lambda})\dot{x}^{\alpha}\dot{x}^{\beta}-\frac{1}{2}g^{\lambda
\kappa}(f_{\lambda,\mu}-f_{\mu,\lambda})\dot{x}^{\mu}-\frac{1}{2}g^{\lambda
\kappa}\phi_{,\lambda}=0,\label{eq19}
\end{equation}
where $V$ is given by equation (\ref{eq14}) and $g_{\lambda\beta}
g^{\lambda \kappa}=\delta^{\kappa}_{\beta}$. Knowing that
$\dot{x}^{4}=-1
,~f_{\lambda}=\hat{g}_{4\lambda},~\phi=\hat{g}_{44}$ and partial
derivatives with respect to $x^{4}$ are zero, we can write
(\ref{eq19}) as
\begin{equation}
\ddot{x}^{\kappa}+\Gamma^{\kappa}_{AB}\dot{x}^{A}\dot{x}^{B}=0,
\end{equation}
where
\begin{equation}
\Gamma^{\kappa}_{AB}=\frac{1}{2}g^{\lambda
\kappa}(\hat{g}_{A\lambda,B}+\hat{g}_{B\lambda,A}-\hat{g}_{AB,\lambda}).
\end{equation}
If we write the Riemann  tensor
\begin{equation}
\hat{R}^{A}_{BCD}=\partial_{C}\Gamma^{A}_{BD}-\partial_{D}\Gamma^{A}_{BC}+
\Gamma^{E}_{BD}\Gamma^{A}_{CE}-\Gamma^{E}_{BC}\Gamma^{A}_{DE},
\end{equation}
we can easily see that $\Gamma^{4}_{AC}=0$. We then have
$\hat{R}{^4}_{BCD}=0$ and
\begin{equation}
\hat{R}^{\alpha}_{BCD}=\partial_{C}\Gamma^{\alpha}_{BD}-
\partial_{D}\Gamma^{\alpha}_{BC}+\Gamma^{\delta}_{BD}\Gamma^{\alpha}_{C
\delta}-\Gamma^{\delta}_{BC}\Gamma^{\alpha}_{D\delta}.
\end{equation}
We also see that $\hat{R} ^{\alpha}_{BCD}$ is a tensor in four
dimensional space which is characterized by $g_{\alpha \lambda}$.
Noting that
\begin{equation}
\Gamma^{\gamma}_{\alpha
\beta}=\frac{1}{2}g^{\lambda\gamma}(\hat{g}_{\alpha
\lambda,\beta}+\hat{g}_{\beta \lambda,\alpha}-\hat{g}_{\alpha
\beta,\lambda}),
\end{equation}
and that $\hat{g}_{\alpha \lambda}=g_{\alpha \lambda}$, we find that
$\hat{R}^{\alpha}_{\beta \gamma \sigma}=R^{\alpha}_{\beta \gamma
\sigma}$ where $R^{\alpha}_{\beta \gamma \sigma}$ is the usual
Riemann tensor. We may now show that
\begin{equation}
\hat{R}_{\alpha \beta \kappa
4}=\frac{1}{2}\nabla_{\kappa}F_{\alpha \beta}.
\end{equation}
The first Bianchi identity is
\begin{eqnarray}
\hat{R}_{\alpha \beta \kappa 4}+\hat{R}_{\kappa \alpha \beta
4}+\hat{R}_{\beta\kappa \alpha 4}=0, \label{eq38}
\end{eqnarray}
leading to
\begin{eqnarray}
\frac{1}{2}\nabla_{\kappa}F_{\alpha
\beta}+\frac{1}{2}\nabla_{\beta}F_{\kappa
\alpha}+\frac{1}{2}\nabla_{\alpha}F_{\beta \kappa}=0, \label{eq39}
\end{eqnarray}
showing that, as expected, it results in the homogeneous Maxwell
equations. It is worth mentioning that in other $5D$ theories, e.g.
Kaluza-Klein, equation (\ref{eq38}) does not lead to equation
(\ref{eq39}). Now, introducing the Ricci tensor as
$\hat{R}_{BD}=\hat{R}^{\alpha}_{B \alpha D}$, we can easily see that
\begin{equation}
\hat{R}_{\beta 4}=\hat{R}^{\alpha}_{\beta \alpha
4}=\frac{1}{2}\nabla_{\alpha}F^{\alpha}_{\beta}.
\end{equation}
The field equations can now be written as follows
\begin{equation}
\hat{G}_{AB}=\hat{R}_{AB}-\frac{1}{2}\hat{g}_{AB}R=8\pi\hat{T}_{AB}\label{eq1}.
\end{equation}
The above equation can be cast into the $4D$ Maxwell and Einstein
equations where $R$ is defined as
\begin{equation}
R=g^{\lambda\alpha}R_{\lambda\alpha},
\hspace{1cm}R_{\lambda\alpha}=\hat{R}_{\lambda\alpha}.
\end{equation}
Let us introduce the five dimensional energy-momentum tensor
according to
\begin{equation}
\hat{T}_{AB}=\left(%
\begin{array}{ccccc}
  T_{\alpha\beta} &   &   &   & \frac{q}{m}\kappa j_{0} \\
    &   &   &   & \frac{q}{m}\kappa j_{1} \\
    &   &   &   & \frac{q}{m}\kappa j_{2} \\
    &   &   &   & \frac{q}{m}\kappa j_{3} \\
  \frac{q}{m}\kappa j_{0} & \frac{q}{m}\kappa j_{1} & \frac{q}{m}\kappa j_{2} & \frac{q}{m}\kappa j_{3} & T_{44} \\
\end{array}%
\right), \label{eq2}
\end{equation}
where $\kappa$ is a coupling constant and $j$ is the current
vector in the $4D$ space-time. The equation of continuity
$j^{\alpha}_{\,\;;\alpha}=0$ now gives us the conservation law. If
$A\rightarrow \alpha$ and $B\rightarrow \beta$, equation
(\ref{eq1}) reduces to the $4D$ Einstein equation since
\begin{equation}
\hat{R}_{\beta \sigma}=R_{\beta \sigma},\hspace{.5cm}
\hat{g}_{\alpha \lambda}=g_{\alpha
\lambda},\hspace{.5cm}\hat{T}_{\beta \sigma}=T_{\beta \sigma}.
\end{equation}
However, if $A\rightarrow \alpha,B=4$, equation $(\ref{eq1})$
reduces to the Maxwell equations with a correction term
\begin{equation}
\hat{G}_{4\lambda}=\hat{R}_{4\lambda}-\frac{1}{2}\hat{g}_{4\lambda}R=8\pi\hat{T}_{4\lambda}=
8\pi\frac{q}{m}\kappa j_{\lambda}, \label{eq50}
\end{equation}
and
\begin{equation}
\nabla_{\alpha}F^{\alpha}_{\lambda}-f_{\lambda}R=16\pi
\frac{q}{m}\kappa j_{\lambda}. \label{eq2}
\end{equation}
Here we have used
$$\hat{R}_{\beta 4}=\frac{1}{2}\nabla_{\alpha}F^{\alpha}_{\beta},~~~~\hat{T}_{4\lambda}=\frac{q}{m}\kappa
j_{\lambda},~~~\mbox{and}~~~~\hat{g}_{4\lambda}=f_{\lambda}.$$

\subsection{Modified Maxwell equations} From (\ref{eq2}) we can
write the modified Maxwell equations as
\begin{equation}
\partial_{\alpha}F^{\alpha \beta}+\Gamma^{\alpha}_{\alpha
\lambda}F^{\lambda \beta}=R f^{\beta}+16\pi \frac{q}{m}\kappa
j^{\beta}, \label{eq3}
\end{equation}
where
\begin{eqnarray}
 F^{\alpha \beta}=\frac{q}{m} \left(
\begin{array}{cccc}
  0 & -\sqrt{\mu_0 \epsilon_0}E_x & -\sqrt{\mu_0 \epsilon_0}E_y & -\sqrt{\mu_0 \epsilon_0}E_z \\
  \sqrt{\mu_0 \epsilon_0}E_x & 0 & -B_z & B_y \\
  \sqrt{\mu_0 \epsilon_0}E_y & B_z & 0 & -B_x \\
  \sqrt{\mu_0 \epsilon_0}E_z & -B_y & B_x & 0 \\
\end{array}
\right),
\end{eqnarray}
and
\begin{eqnarray}
f^{\beta}&=&\frac{q}{m}
\left(\sqrt{\mu_0 \epsilon_0}\phi \,\,\, A_x \,\,\, A_y \,\,\, A_z \right), \nonumber \\
j^{\beta}&=&\left( \rho \,\,\, \sqrt{\mu_0 \epsilon_0}j_x \,\,\,
\sqrt{\mu_0 \epsilon_0}j_y \,\,\, \sqrt{\mu_0 \epsilon_0}j_z
\right). \nonumber
\end{eqnarray}

\section{Origin of galactic magnetic fields}
We are now ready to use the results obtained above to show how the
galactic magnetic fields are produced.  Let us first suppose that
our galaxy is disk-like, having a thickness $2d$, radius $l$ which
is assumed to be large. The $z$ axis is taken as being perpendicular
to the plane of the disk and the origin of coordinates is on the
geometric center of the disk having a homogenous mass density
$\rho$. Now suppose that a plane wave is incident on the disk and
passes through the galaxy at $y=0$ and $z=-d$. This wave can be
represented by
\begin{equation}
f_{in}^{\beta}=\left(%
\begin{array}{cccc}
  0 & 0 & 0 & A' \cos(k'_2 y+k'_3(z+d) -\omega t) \\
\end{array}%
\right),\label{eq4}
\end{equation}
\begin{equation}
f_{out}^{\beta}=\left(%
\begin{array}{cccc}
  0 & 0 & 0 & A \cos(k_2 y+k_3(z+d) -\omega t) \\
\end{array}%
\right).\label{eq5}
\end{equation}
\subsection{Weak field approximation}
We may now use the weak field approximation and write the metric as
\begin{equation}
g_{m n}=\eta_{m n}+h_{m n},
\end{equation}
where $\eta_{m n}$ is the Minkowski metric and $h_{m n}\ll 1 $. We
then have
\begin{equation}
\Gamma^{\sigma}_{\mu \nu}=\frac{1}{2}\eta^{\sigma
\rho}(\partial_{\nu}h_{\rho \mu}+\partial_{\mu}h_{\rho
\nu}-\partial_{\rho}h_{\mu \nu}),
\end{equation}
\begin{equation}
R=-\Box (\eta^{\rho \sigma}h_{\rho
\sigma})+\partial_{\rho}\partial_{\mu}(\eta^{\mu \nu}\eta^{\rho
\lambda}h_{\nu \lambda}),
\end{equation}
with
\begin{equation}
h_{00}=h_{11}=h_{22}=h_{33}=h=\frac{-2\phi}{c^2},
\end{equation}
where $c$ is the speed of light and $\phi$ is the Newtonian
potential. Newtonian potential for the galaxy  model described above
can be easily written as follows
\begin{equation}
\begin{array}{cc}
  ~~~~\phi_{in}=2\pi G \rho z^2,~~~~~~~~~~~~~~~ & ~~~~~~~~~~~~~~\phi_{out}=4\pi G d \rho z-2\pi G \rho d^2, \\
\end{array}
\end{equation}
where the indices refer to the inner and outer galaxy regions. Then
we have
\begin{eqnarray}
  h_{in}&=&\frac{-4 \pi G \rho z^2}{c^2}, \hspace{28mm} h_{out}=\frac{-8 \pi G \rho d z}{c^2}+
  \frac{4\pi G \rho d^2}{c^2},\nonumber \\
R_{in}&=&\frac{8\pi G \rho}{c^2}, \hspace{34mm} R_{out}=0,\nonumber \\
  \\
  \Gamma_{in ~\alpha 0}^{\alpha}&=&\Gamma_{in~ \alpha 1}^{\alpha}=\Gamma_{in~ \alpha 2}^{\alpha}=0, \hspace{10mm}
  \Gamma_{out ~\alpha 0}^{\alpha}=\Gamma_{out~ \alpha 1}^{\alpha}=\Gamma_{out~ \alpha 2}^{\alpha}=0,\nonumber \\
  \Gamma_{in~\alpha 3}^{\alpha}&=&\frac{-8\pi G \rho z}{c^2}, \hspace{28mm}
  \Gamma_{out~ \alpha 3}^{\alpha}=\frac{-8\pi G d \rho}{c^2}.
  \nonumber
\end{eqnarray}
for more detailed description see \cite{20}. If $k_3$ is arbitrarily
small and $\omega$ tends to zero, equations (\ref{eq4}) and
(\ref{eq5}), to a good approximation, are the solutions of equation
(\ref{eq3}) and we have
\begin{eqnarray}
k_{2}^2=\frac{\frac{\omega^2}{c^2}(1+h_{out})}{(1-h_{out})},\hspace{20mm}
k_{2} ^{'2} = \frac{\frac{\omega^2}{c^2}(1-h_{in}) + R}{1-h_{in}},
\end{eqnarray}
where the terms containing higher orders of $\frac{G \rho}{c^2}$
have been neglected.
\subsection{Continuity of currents}
Since currents must satisfy continuity equations, so does our new
current. This means that we must have
\begin{equation}
\nabla_{\beta}(Rf^{\beta})=0.
\end{equation}
By neglecting higher orders of $\frac{G \rho}{c^2}$ we are led to
\begin{equation}
R\partial_{\beta}f^{\beta}=0.
\end{equation}
Since $R$ is very small, this equation is approximately satisfied.
\section{Boundary conditions}
Since the metric should be continuous across the the boundaries
$y=0$ and $ z=-d$, we find from equations (\ref{eq4}) and
(\ref{eq5}) that $A=A'$. If our goal is to find the ratio of the
field strengths, we can neglect the terms containing $\frac{G
\rho}{c^2}$ in the Faraday tensor. Therefore, from equations
(\ref{eq4}) and (\ref{eq5}) we find
\begin{equation}
\begin{array}{cc}
  F^{0 3}_{in} \simeq \frac{A' \omega}{c} \sin(k'_{2} y + k'_{3} (z+d)-\omega t), & F^{0 3}_{out}\simeq \frac{A \omega}{c}
  \sin(k_{2} y + k_{3} (z+d)-\omega t), \\
  \\
  F^{2 3}_{in} \simeq A' k'_{2} \sin(k'_{2} y + k'_{3} (z+d)-\omega t), & F^{2 3}_{out} \simeq A k_{2}
  \sin(k_{2} y + k_{3} (z+d)-\omega t). \\
\end{array}
\end{equation}
Now it is clear that
\begin{equation}
\begin{array}{ccc}
  h_{in}=h_{out}|_{boundary}, & \frac{E_{in_z}}{E_{out_z}}=1|_{boundary}, & \frac{B_{in_x}}{B_{out_x}}=
  \frac{k'_{2}}{k_{2}}|_{boundary}. \\
\end{array}
\end{equation}
We finally obtain the desired relation
\begin{equation}
\frac{B_{in_x}}{B_{out_x}} \simeq \sqrt{1 + \frac{R c^2}{\omega
^2}}.
\end{equation}
Now for the average density of a galaxy \cite{21}, $\bar{\rho} \sim
10^{-21}kg m^{-3}$, the seed magnetic field which is of order of
$\sim 10^{-20}G$ would be amplified to $\sim 10^{-6}G$, if we take
$\omega\sim 10^{-29}Hz $. This indicates that galactic magnetic
fields are almost static.
\section{Conclusions}
In this paper we have employed a theory resulting from a fresh look
at the unification of gravity and electromagnetism, developed
previously, to explain one of the most intriguing phenomena at the
galactic scales, namely the origins of the unexpectedly large
magnetic fields which one measures within galaxies. We showed that
if an electromagnetic field is incident upon a galaxy parallel to
its plane, it would be amplified because of the coupling of the
vector potential to the scalar curvature produced by the mass
distribution in the galaxy. We have also shown that the
amplification occurs for magnetic fields which are almost static, an
assumption common in such investigations.

\end{document}